\documentclass[runningheads]{llncs}

\usepackage{amsmath}
\usepackage{subcaption}
\usepackage{graphicx}
\usepackage{amsfonts}
\usepackage{multirow}
\usepackage{url}
\usepackage{listings}
\usepackage{tikz}
\usetikzlibrary{arrows.meta}
\usetikzlibrary{matrix}
\usetikzlibrary{positioning}
\usetikzlibrary{shapes.geometric}
\tikzstyle{every node}=[font=\footnotesize]
\tikzstyle{etichetta}=[]
\tikzstyle{square}=[draw,outer sep=0pt,inner sep=-1.3em,regular polygon,regular polygon sides=4,,minimum size=13.5mm]
\tikzstyle{square2}=[square,dotted,black!50]
\pgfmathsetmacro{\ra}{0.95}
\usetikzlibrary{decorations.pathreplacing,calc}

\usepackage{listings}
\lstdefinelanguage{json}{
    showspaces=false,
    showtabs=false,
    breaklines=false,
    breakatwhitespace=false,
    basicstyle=\ttfamily\footnotesize,
    upquote=true,
    morestring=[b]",
    literate={♭}{$\flat$}1
}
\usepackage{algorithm}
\usepackage{algpseudocode}

\algnewcommand\algorithmicswitch{\textbf{switch}}
\algnewcommand\algorithmiccase{\textbf{case}}
\algnewcommand\algorithmicdefault{\textbf{default}}
\algnewcommand\algorithmicassert{\texttt{assert}}
\algnewcommand\Assert[1]{\State \algorithmicassert(#1)}%
\algdef{SE}[SWITCH]{Switch}{EndSwitch}[1]{\algorithmicswitch\ #1\ \algorithmicdo}{\algorithmicend\ \algorithmicswitch}%
\algdef{SE}[CASE]{Case}{EndCase}[1]{\algorithmiccase\ #1}{\algorithmicend\ \algorithmiccase}%
\algdef{SE}[DEFAULT]{Default}{EndCase}[1]{\algorithmicdefault\ #1}{\algorithmicend\ \algorithmiccase}%
\algtext*{EndSwitch}%
\algtext*{EndCase}%
\algnewcommand\algorithmicforeach{\textbf{for each}}
\algdef{S}[FOR]{ForEach}[1]{\algorithmicforeach\ #1\ \algorithmicdo}

\begin{document}

\title{A Dual-CRDT Architecture for Decentralized Trust Governance and Evolution}

\titlerunning{Dual-CRDT Trust Governance and Evolution}

\author{Amos Brocco\inst{1}}
\authorrunning{A. Brocco}
\institute{University of Applied Sciences and Arts of Southern Switzerland, Lugano, Switzerland
\email{amos.brocco@supsi.ch}}

\maketitle 

\begin{abstract}
While CRDTs provide decentralized replication and eventual
consistency, Byzantine-resilient deployments require
mechanisms for deciding which updates should be trusted and
therefore contribute to the reconstructed state. In practice,
the trust relationships underlying these decisions may
evolve over time as participants join or leave, identities
change, and governance rules are revised. However, the
information used to make trust decisions is typically
managed outside the replicated state itself.
This paper introduces a dual-CRDT architecture composed of a
\emph{Trust CRDT} and a \emph{Data CRDT}. The Trust CRDT
stores and evolves governance information, while the Data
CRDT is reconstructed according to the trust configuration
derived from the Trust CRDT. Governance therefore becomes
replicated state rather than an externally managed artifact.
Building upon deterministic reconstruction and Byzantine
trust filtering, the proposed model allows trust
relationships and governance rules to evolve through ordinary
CRDT updates. 
The resulting architecture provides a recursive
governance model in which governance rules determine their
own future evolution while simultaneously
governing application data.
The approach is implemented as a prototype on top of Melda
and melda-sec and should be viewed as an initial exploration
of decentralized trust governance and evolution for
Byzantine-resilient CRDT systems.
\end{abstract}
\section{Introduction}

Conflict-free Replicated Data Types (CRDTs)~\cite{b1,b10,b32} provide Strong Eventual Consistency (SEC) without coordination, allowing replicas to evolve independently while guaranteeing convergence when they observe the same set of updates~\cite{b0}. Delta-state CRDTs ($\delta$-CRDTs)~\cite{b3,b14} further improve efficiency by propagating compact, idempotent deltas rather than complete states.

In open and decentralized environments, however, data consistency is only part of the problem. Replicas must also determine which updates should be trusted. Practical systems therefore require trust configurations defining trusted identities, authorization policies, endorsement requirements, and revocation mechanisms. Traditionally, such trust configurations are managed externally to the replicated data itself. Public keys, authorization rules, and security policies are typically distributed out-of-band through administrative procedures or centralized infrastructure. While sufficient in static environments, this approach becomes problematic in decentralized systems where trust relationships evolve over time. Participants may join or leave, authorization policies may change, cryptographic keys may be rotated, and previously accepted content may later need to be revoked.

In our previous work, we introduced deterministic reconstruction as a mechanism for achieving Byzantine-resilient convergence without requiring global agreement on update validity~\cite{b35}. We subsequently extended this model with fine-grained trust management, decoupling identity-based trust from content-based trust and introducing whitelist and blacklist mechanisms for selective post-compromise handling~\cite{b36}. Both approaches assume the existence of a trust configuration used during reconstruction, but leave open the question of how such a configuration is established, distributed, and evolved in a decentralized system.

In this paper, we introduce a dual-CRDT architecture composed of a \emph{Trust CRDT} and a \emph{Data CRDT}. The Trust CRDT maintains the current trust configuration, while the Data CRDT stores application data. State reconstruction of the Data CRDT is parameterized by the trust configuration obtained from the reconstructed state of the Trust CRDT. As a result, trust becomes a replicated and evolvable part of the system state rather than an externally managed configuration. Participants can therefore collaboratively modify not only application data, but also the governance rules determining which future updates are considered acceptable.

The proposed architecture should be viewed as an early exploration of decentralized trust governance for CRDT systems. The current prototype is implemented on top of Melda\footnote{\url{https://github.com/slashdotted/libmelda}} and the melda-sec\footnote{\url{https://github.com/slashdotted/libmelda-sec}} trust layer, which provide a convenient foundation for experimenting with trust evolution, endorsement-based governance, and deterministic reconstruction.

At this stage, this proposition should be considered work in progress rather than a fully validated system. Nevertheless, the model is intentionally described at an abstract level and is not inherently tied to Melda. An important direction for future work is to investigate its applicability to other CRDT families and reconstruction models, and to identify the conditions under which a replicated trust configuration can safely govern the evolution of replicated application state.

\section{Background}
This work builds upon deterministic reconstruction
for Byzantine-resilient convergence~\cite{b35}
and fine-grained trust-aware filtering~\cite{b36}.
Both approaches assume the existence of a trust
configuration used during reconstruction.
This paper addresses how such a configuration
can itself be represented, distributed,
and evolved within a decentralized CRDT system.

\section{The Trust Evolution Problem}
The trust model presented in our previous work~\cite{b36} assumes the existence of a trust configuration defining:
\begin{itemize}
\item trusted identities;
\item authorization policies;
\item endorsement requirements;
\item content-level trust decisions.
\end{itemize}

Such a configuration determines which updates are allowed to contribute to state reconstruction.
A natural question then arises: \emph{how should this trust configuration itself be managed?}
A static trust configuration may be sufficient for short-lived or centrally administered deployments. However, in long-lived decentralized systems, trust relationships are rarely static. Participants may join or leave the system.
Cryptographic keys may be rotated. Authorization policies may evolve. Previously accepted content may later require revocation. Consequently, trust becomes a dynamic property rather than a one-time configuration choice.

\paragraph{Example} Consider a collaborative application governed by a trust
configuration in which Michael, David, and Lukas act as
trustees. The corresponding data policy allows only Michael
and David to modify application data.
Over time, the governance of the system may evolve.
For example, the participants may decide that Anna should
become a trustee, replacing Lukas. At the same time, the
authorization policy may be changed so that Anna and Lukas
are allowed to modify application data, while Michael and
David are no longer authorized.
This change does not merely affect the future behavior
of the system. It also affects the rules used to evaluate
future updates. Consequently, the trust configuration
itself becomes part of the evolving state of the system.

A more subtle challenge arises from the fact that trust
configurations participate in their own evolution.
As an example, suppose that a new trust configuration is accepted.
The configuration may alter the set of trusted identities,
change endorsement requirements, or modify authorization
policies. If these new rules were applied retroactively,
they might invalidate historical updates that were required
to establish the new configuration in the first place.

This creates a circular dependency: the trust configuration determines which updates are trusted,
while updates are themselves used to evolve the trust
configuration. Therefore, trust evolution cannot be reduced to simply
replacing one configuration with another. The system must
provide a mechanism allowing trust to evolve without
destroying the historical information upon which that
evolution depends.

To address this problem, we propose representing trust as
replicated state. Rather than distributing trust
configurations through external administrative procedures,
they are maintained by a dedicated Trust CRDT whose state
evolves through the same decentralized mechanisms used for
application data.

\section{Trust Mechanisms}
\label{sec:tarec}

Before introducing the proposed dual-CRDT architecture, it is
necessary to describe the trust mechanisms on which it is
built. These mechanisms are provided by \emph{melda-sec} and can be
used independently of the trust-evolution model presented in
this paper. They define how trust decisions are evaluated
during reconstruction, how updates are approved, and how
authorization rules are enforced. In the proposed
architecture, the same mechanisms are applied both to
governance updates stored in the Trust CRDT and to
application updates stored in the Data CRDT.

The core component of the framework is the
\emph{TrustAdapter}, a composable adapter that extends a
Melda CRDT with trust-aware reconstruction capabilities.
Rather than preventing updates from being created,
propagated, or stored, the TrustAdapter determines which
updates are allowed to contribute to the reconstructed state.

Conceptually, every update is treated as a proposal. A
proposal becomes part of the reconstructed state only if it
satisfies the active trust configuration. This approach
separates update dissemination from trust evaluation:
updates may be exchanged freely between participants, while
trust decisions are applied deterministically during state
reconstruction.

\subsection{Trusted Identities}
The trust model is based on cryptographic identities.
Each participant is associated with a public/private key
pair. The private key remains under the control of the
participant, while the public key serves as a stable
identifier within the trust system.

TrustAdapter maintains a set of trusted public keys together
with optional metadata, such as roles and authorization
attributes. The active trust configuration determines which
public keys are considered trusted at any given time.
This approach decouples trust from network addresses,
replica identifiers, or deployment-specific information.
Trust relationships are instead expressed through
cryptographic identities that can be added, removed, or
reassigned as the governance state evolves.

Trusted identities provide the foundation upon which the
remaining trust mechanisms are built. In particular, they
allow participants to express approval decisions and to
enforce authorization policies independently of the
underlying replication infrastructure.

\subsection{Endorsements}
TrustAdapter relies on cryptographic \emph{endorsements} to
express trust decisions. An endorsement is a digital signature associating a trusted
identity with a specific delta generated by a Melda commit.
By issuing an endorsement, a participant explicitly approves
the inclusion of that delta during state reconstruction.

This model separates delta creation from delta approval. Any
participant may generate, store, and disseminate deltas,
while trust is expressed through endorsements provided by
trusted identities. Consequently, reconstruction depends not
only on who produced a delta, but also on who is willing to
approve it. The same endorsement mechanism is used throughout the
architecture. In particular, governance deltas stored in the
Trust CRDT are treated in exactly the same way as application
deltas stored in the Data CRDT and are evaluated according to
the currently active endorsement policy. As a result, the
same trust model can be applied both to application data and
to the evolution of the governance state itself.

\subsubsection{Endorsement Modes}

Different systems may require different trust assumptions.
TrustAdapter therefore supports multiple endorsement modes,
allowing reconstruction policies to range from permissive
deployment scenarios to decentralized governance models.
The current implementation provides a small set of commonly
used endorsement policies. The framework is intentionally
extensible, and additional endorsement strategies may be
introduced in future versions.

\paragraph{Permissive mode}
In permissive mode, delta blocks are accepted regardless of
their trust status. This mode is primarily intended for
debugging, configuration validation, and fault-injection
experiments.

\paragraph{Single endorsement}
In single-endorsement mode, a delta block is accepted once it
has been endorsed by at least one trusted and authorized
identity.
This mode is suitable for deployments in which approval by a
single maintainer or administrator is sufficient.

\paragraph{Majority endorsement}

In majority-endorsement mode, a delta block is accepted only
if a majority of the currently trusted identities endorse it.
For example, a system with five trustees requires at least
three endorsements before a delta contributes to the
reconstructed state.
This mode enables decentralized governance by distributing
approval authority among multiple trustees rather than
concentrating it in a single participant. It is particularly
relevant for trust evolution, where governance deltas may
themselves require approval by the trustees defined in the
currently active trust configuration.

\paragraph{Unanimous endorsement}

In unanimous-endorsement mode, a delta block is accepted only
if all currently trusted identities endorse it.
This mode provides the strongest approval guarantees and is
appropriate for highly sensitive governance actions where
every trustee must explicitly agree before a change becomes
effective. The stricter approval requirement reduces the risk
of controversial governance changes, at the cost of
requiring full participation from all trusted identities.

\subsection{Authorization Policies}
Endorsements alone are not sufficient. Trusted identities
may have different responsibilities within the system.
TrustAdapter therefore includes a policy engine that
evaluates authorization rules during reconstruction.
Authorization decisions are based on the object identifiers
declared by delta blocks. Each delta explicitly specifies
which objects it creates or modifies, allowing the policy
engine to determine which parts of the application state are
affected by that delta.

Policy evaluation is performed using these object
identifiers together with the identity and role of the
authoring participant. Policies may target:

\begin{itemize}
\item specific public keys;
\item roles associated with trusted identities;
\item subsets of application objects identified by their
      object identifiers.
\end{itemize}

The default policy is \emph{deny all}. A delta is considered
authorized only if at least one policy rule explicitly grants
permission for the corresponding identity or role to affect
the object identifiers declared by the delta.
By default, TrustAdapter operates in \emph{strict write
mode}. Under this mode, policy evaluation is performed before
a delta is committed. If the identity associated with the
local signing key is not authorized to create or modify the
objects declared by the delta, the commit operation is
rejected and the delta is not stored.
Strict write mode can be disabled when testing or evaluating
the behavior of the system. In this configuration,
unauthorized deltas may still be generated and disseminated,
allowing experiments involving misconfigured or malicious
participants.

Policy enforcement also occurs during reconstruction.
Regardless of how a delta was produced or propagated, a delta
that declares modifications violating the active policy is
not accepted by the TrustAdapter and therefore does not
participate in deterministic reconstruction.
As a result, policies remain independent of the internal
representation or semantic content of application data.
Authorization rules determine which object identifiers may
be affected by a given identity or role, rather than how the
corresponding objects are modified.
This allows governance decisions to distinguish between
trust and authority. A participant may be trusted by the
system, while authorization policies determine which object
categories that participant may create or modify, and for
which object categories the participant's endorsements are
considered valid.

\subsection{Content-Level Trust}

TrustAdapter also supports explicit content-level trust
decisions through whitelist and blacklist mechanisms.
Whitelisted delta blocks are always accepted during
reconstruction, regardless of endorsement requirements,
authorization policies, or other trust constraints.
Conversely, blacklisted delta blocks are always excluded,
even if they would otherwise satisfy the active trust
configuration.

These mechanisms provide a fine-grained trust layer that
operates at the level of individual deltas rather than
identities or roles. As a result, trust in a participant can
be distinguished from trust in the content produced by that
participant. This distinction is particularly useful during trust
evolution and post-compromise handling. For example, a delta
may later be discovered to be incorrect, malicious, or
undesirable. Rather than revoking the identity that produced
or endorsed it, a governance update may explicitly blacklist
that specific delta while preserving all other contributions
associated with the same identity.
Similarly, whitelist entries can be used to preserve
historical deltas that should remain part of the reconstructed
state regardless of future governance changes.

\section{The Trust CRDT}
The mechanisms described in the previous section determine
how a trust configuration is evaluated. They define how
trusted identities, endorsements, authorization policies,
and content-level trust decisions influence state
reconstruction. However, they do not address how a trust
configuration is created, distributed, or evolved.

We address this problem by representing trust itself as
replicated state maintained by a dedicated \emph{Trust CRDT}.
Rather than relying on externally managed configuration
files, administrative procedures, or centralized trust
authorities, governance information becomes part of the
replicated state and evolves through ordinary CRDT updates.

The Trust CRDT stores a governance document containing the
information required to derive the active trust
configuration, including trusted identities, roles,
authorization policies, endorsement settings, and
content-level trust rules.
Replicas reconstruct the Trust CRDT using the same
deterministic reconstruction mechanisms applied to ordinary
application data. The resulting governance document is then
used to derive the active trust configuration for the
system. As a result, trust evolves through the same decentralized
replication mechanisms used for application data, while
remaining compatible with deterministic reconstruction.

\subsection{Governance Updates}
Changes to governance are represented as ordinary updates of
the Trust CRDT. Rather than being applied through external
administrative procedures, governance changes become part of
the replicated history of the system. Examples of governance updates include:

\begin{itemize}
\item adding or removing trusted identities;
\item assigning or revoking roles;
\item changing authorization policies;
\item changing endorsement modes;
\item introducing whitelist entries;
\item introducing blacklist entries.
\end{itemize}

\subsection{Governance Approval}
Governance updates do not automatically become active.
Instead, changes to the governance document are evaluated
using the trust configuration currently reconstructed from
the Trust CRDT. A governance update therefore becomes
effective only after satisfying the endorsement and
authorization requirements defined by the active governance
state. For example, if the current governance configuration
requires majority endorsement, a proposal modifying the set
of trustees must first obtain approval from a majority of
the trustees defined by that same configuration.
This allows governance changes to be collectively evaluated
by the identities currently entrusted with governing the
system.

\subsection{Recursive Governance}
The Trust CRDT introduces a recursive relationship between
trust and governance. The trust configuration reconstructed from the Trust CRDT is
used to evaluate new governance updates. Accepted governance
updates contribute to the future state of the Trust CRDT and
may therefore modify the trust configuration reconstructed
later. As a result, the rules governing trust evolution are
themselves subject to evolution. The Trust CRDT therefore
provides a self-governing mechanism in which governance
changes are evaluated according to the governance state that
is active when those changes are proposed.

\section{The Dual-CRDT Governance Architecture}
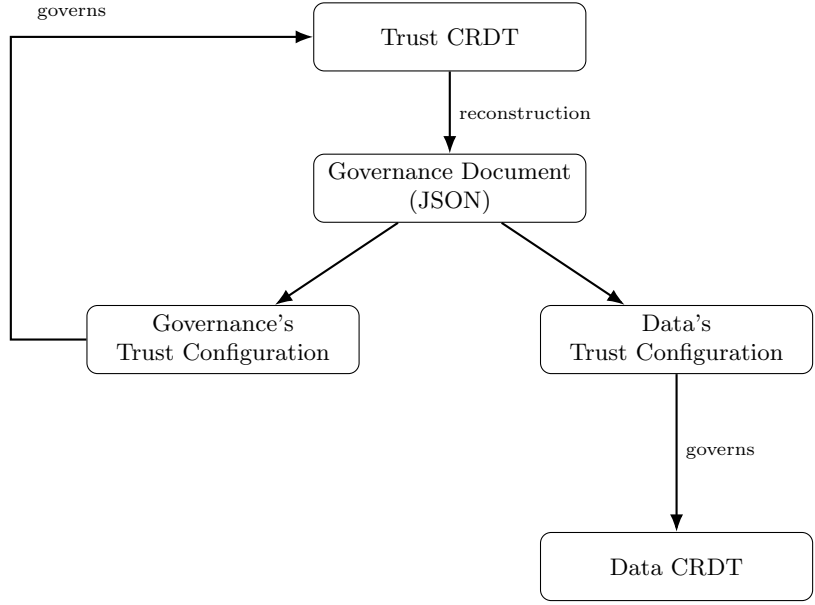
\begin{figure}[t]
\centering

\begin{tikzpicture}[
box/.style={
draw,
rounded corners,
align=center,
minimum width=3.6cm,
minimum height=0.9cm
},
arrow/.style={
-Latex,
thick
},
every node/.style={font=\small}
]

\node[box] (trustcrdt) at (0,4)
{Trust CRDT};

\node[box] (govdoc) at (0,2)
{Governance Document\\(JSON)};

\node[box] (trustcfg) at (-3,0)
{Governance's\\Trust Configuration};

\node[box] (datacfg) at (3,0)
{Data's\\Trust Configuration};

\node[box] (datacrdt) at (3,-3)
{Data CRDT};

\draw[arrow] (trustcrdt) --
node[right,font=\scriptsize]{reconstruction}
(govdoc);

\draw[arrow] (govdoc) -- (trustcfg);
\draw[arrow] (govdoc) -- (datacfg);

\draw[arrow]
(trustcfg.west)
-- ++(-1,0)
-- ++(0,4.0)
-- (trustcrdt.west);

\node at (-5.0,4.3)
{\scriptsize governs};

\draw[arrow]
(datacfg)
--
node[right,font=\scriptsize]{governs}
(datacrdt);

\end{tikzpicture}

\caption{
Recursive governance in the proposed dual-CRDT architecture.
The Trust CRDT reconstructs a governance document containing
both the trust policy governing its own future evolution and
the trust policy governing the Data CRDT. Changes to the
governance document therefore influence both the Trust CRDT
and the Data CRDT.
}
\label{fig:dualcrdt}
\end{figure}

The proposed architecture separates governance from
application data while representing both as replicated state. Instead of embedding trust information directly into the application CRDT, the system is composed of two interacting
replicated data structures:

\begin{itemize}
\item a \emph{Trust CRDT}, responsible for storing and
evolving governance information;
\item a \emph{Data CRDT}, responsible for storing
application data.
\end{itemize}

As illustrated in Figure~\ref{fig:dualcrdt}, replicas first reconstruct the Trust CRDT and obtain a governance document. This document contains the information required to derive two
distinct trust configurations: one governing the evolution of the Trust CRDT itself, and one governing the Data CRDT. At the same time, the governance document defines how application data stored in the Data CRDT is evaluated and reconstructed.
State reconstruction therefore proceeds in two stages:

\begin{enumerate}
\item the Trust CRDT is reconstructed and a governance
      document is derived;
\item the resulting trust configuration is applied during
      reconstruction of the Data CRDT.
\end{enumerate}

The model is recursive. Governance rules evolve through
updates stored in the Trust CRDT, while those same rules
determine which future governance updates may become
effective.

To bootstrap the process, replicas begin with an initial
trust configuration, referred to as the \emph{bootstrap
configuration}. Using this configuration, replicas
reconstruct the Trust CRDT and derive a governance document.
The resulting governance document is then used to derive a
new trust configuration, which is applied during a subsequent
reconstruction pass.

This process is repeated until no additional governance
changes become active. Since governance updates are finite
within a finite Trust CRDT history, reconstruction
eventually reaches a fixed point. At that point, replicas
obtain the most recent trust configuration derivable from the
available Trust CRDT history and use it to reconstruct the
Data CRDT.

Consequently, the active governance state emerges from the
replicated history itself rather than from an externally
distributed configuration file, while governance and
application data remain logically separate.

\subsection{Governance-Driven Reconstruction}
The key consequence of the dual-CRDT architecture is that changes to governance may alter the reconstructed state of the Data CRDT without modifying the underlying data type.
When the governance document evolves, replicas derive a new trust configuration and apply it during reconstruction of the Data CRDT. Consequently, the set of accepted deltas may change over time as trust relationships, authorization policies, and endorsement requirements evolve.
This allows the system to adapt to changing governance requirements without requiring centralized administration or modifications to the semantics of the underlying CRDT.

\paragraph{Example}
Consider a system in which Michael, David, and Lukas act as
trustees. The Trust CRDT requires majority endorsement for
governance changes. The active data policy allows only
Michael and David to modify application data.
Under this configuration, deltas proposed by Michael and
David contribute to the reconstructed state, while deltas
produced by other participants are ignored.
Suppose that the trustees later approve a new trust
configuration replacing Lukas with Anna and introducing a new
authorization policy in which Anna and Lukas are allowed to
modify application data.
Once the new configuration receives the required
endorsements, all replicas reconstruct the Trust CRDT and
obtain the updated governance state. Subsequent
reconstruction of the Data CRDT uses the newly derived trust
configuration.
As a result, deltas produced by Anna and Lukas become
eligible for inclusion, while new deltas produced by Michael
and David are no longer authorized.

Recursive governance introduces an additional challenge.
A newly accepted trust configuration may invalidate
governance deltas that were required to establish that same
configuration. To preserve the reconstructability of the
governance history, such historical governance deltas may be
explicitly whitelisted. This ensures that previously accepted
governance decisions remain part of the reconstructed state
even when the active trust configuration evolves beyond the
rules that originally accepted them.
The application data therefore evolves not only because new
deltas are generated, but also because the governance rules
used to interpret those deltas evolve over time.

\subsection{Content Revocation}
Trust evolution is not limited to identities and policies. The trust configuration may also contain explicit content-level decisions. Continuing the previous example, suppose that a previously accepted delta created by David is later discovered to be incorrect. Rather than revoking David's identity or removing him from the set of trusted participants, a new governance update can explicitly blacklist the corresponding delta.
Once the updated trust configuration becomes active, replicas reconstruct the Data CRDT while excluding the blacklisted content. Importantly, David remains trusted and may continue to participate in future governance and data updates.
This distinction between trust in identities and trust in individual deltas enables fine-grained post-compromise and post-fault handling without requiring retroactive exclusion
of all contributions produced by a participant.

\subsection{Governance History Forks}
\label{sec:trust-forks}
The Trust CRDT supports governance reconfiguration through a sequence of endorsed trust updates. In an asynchronous distributed system, however, a trustee set that was valid at a given point in history may generate multiple concurrent governance evolutions originating from the same trusted configuration. Since all such updates are authorized according to the rules in effect at their respective parent states, they are indistinguishable from the protocol's perspective. This behavior does not compromise the convergence properties of the Dual-CRDT model. Given the same set of delta blocks, all replicas deterministically reconstruct the same Trust CRDT state and, consequently, the same interpretation of the associated Data CRDT. However, the protocol cannot determine which concurrent governance evolution represents the intended succession of authority, since no global ordering, consensus layer, or external finality mechanism is assumed. To support governance auditing, implementations may expose the set of fork points encountered during Trust CRDT reconstruction. Such information enables applications to detect competing governance histories and apply domain-specific policies, such as warning operators, triggering manual review procedures, migrating to a new Trust CRDT, or enforcing additional trust requirements. The existence of governance history forks should therefore be regarded as an application-level trust concern rather than a protocol-level consistency issue. The Dual-CRDT model guarantees deterministic reconstruction and convergence, while the interpretation of governance forks is intentionally delegated to the application layer.

A consequence of this design is that historical trustee sets retain the ability to produce alternative governance histories that are valid with respect to their own configuration. Since generating such histories carries no intrinsic protocol cost, the model does not attempt to prevent their creation. Instead, it provides deterministic reconstruction together with fork-detection mechanisms, allowing applications to evaluate whether the observed governance history satisfies their trust assumptions.

\section{Discussion}
The proposed architecture does not introduce a new
convergence model. Both the Trust CRDT and the Data CRDT are
ordinary replicated data structures and therefore inherit the
convergence properties of the underlying deterministic
reconstruction framework.
In particular, omission and equivocation~\cite{b24} affect
the Trust CRDT in the same way they affect any other CRDT
instance. When replicas observe different subsets of
governance deltas, they may temporarily reconstruct
different governance states and therefore derive different
trust configurations.

As a consequence, replicas may initially make different
trust decisions regarding deltas stored in the Data CRDT. For
example, a replica that has not yet observed a governance
update may continue operating under an older trust
configuration. Under that configuration, deltas endorsed by
newly authorized identities may be ignored, while replicas
already observing the updated governance state may accept
them. Similarly, equivocation on governance deltas does not
introduce additional challenges beyond those already present
in the underlying deterministic reconstruction model.
Conflicting governance proposals are treated like any other
deltas and must still satisfy the endorsement and
authorization requirements defined by the active trust
configuration before they can affect the reconstructed
governance state.

These effects are temporary and arise solely from replicas
having different information available at reconstruction
time. Once replicas observe the same set of governance
deltas, deterministic reconstruction guarantees convergence
towards the same governance document and therefore towards
the same trust configuration.
Subsequently, once replicas also observe the same set of
application deltas, they evaluate those deltas under the same
trust configuration and derive the same reconstructed
application state.

Consequently, the proposed dual-CRDT architecture does not
require additional convergence assumptions beyond those
already required by the underlying deterministic
reconstruction model. The architecture extends the scope of
replicated state from application data to governance itself,
while preserving the convergence guarantees of the underlying
CRDT framework.

\section{Conclusion}
Byzantine-resilient CRDT systems typically assume the existence of a
trust configuration defining trusted identities,
authorization policies, endorsement requirements, and
content-level trust decisions. While such configurations are
commonly treated as external administrative artifacts, this
approach becomes increasingly problematic in decentralized
systems where trust relationships evolve over time.

In this paper, we introduced a dual-CRDT architecture that
treats governance itself as replicated state. A dedicated
Trust CRDT maintains a governance document describing the
active trust configuration, while a separate Data CRDT stores
application data. Through deterministic reconstruction, the
Trust CRDT produces the trust configuration used both to
govern its own future evolution and to evaluate application
deltas stored in the Data CRDT.

The resulting model enables decentralized trust evolution
without requiring centralized authorities or out-of-band
distribution of trust policies. Changes to identities,
authorization rules, endorsement policies, and content-level
trust decisions become ordinary replicated updates that are
themselves subject to governance. As a result, trust becomes
an evolving component of the replicated state rather than a
static prerequisite of the system.

The proposed architecture preserves the convergence
properties of the underlying deterministic reconstruction
framework while extending replication from application data
to governance itself. Although the current implementation is
built on top of Melda and melda-sec, the model is intended as
a more general approach to decentralized trust governance.

This work should be considered an initial exploration rather
than a complete solution. Future work will focus on
formalizing the recursive governance model, characterizing
the conditions under which it preserves convergence and
deterministic reconstruction, and evaluating its resilience
under adversarial conditions.
Particular attention will be devoted to Byzantine scenarios
involving equivocation, omission, malicious governance
proposals, key compromise, and attempts to manipulate trust
evolution through conflicting governance histories (as discussed in Section \ref{sec:trust-forks}). Future
research will also investigate the applicability of the model
to other CRDT families and reconstruction frameworks.
More generally, we believe that representing trust as
replicated state opens a promising direction for the design
of self-governing decentralized systems.

\bibliographystyle{splncs04}
\bibliography{dual-crdt}
\end{document}